# Metal-insulator transition upon heating and negative-differential-resistive-switching induced by self-heating in $BaCo_{0.9}Ni_{0.1}S_{1.8}$

B. Fisher, J. Genossar, K. B. Chashka, L. Patlagan and G. M. Reisner

*Physics Department, Technion, Haifa, 32000 Israel*

The layered compound $BaCo_{1-x}Ni_xS_{2-y}$ ($0.05<x<0.2$ and $0.05<y<0.2$) exhibits an unusual first-order structural and electronic phase transition from a low-T monoclinic paramagnetic metal to a high-T tetragonal antiferromagnetic insulator around 200 K with huge hysteresis (~ 40 K) and large volume change (~0.01). Here we report on unusual voltage-controlled resistive switching followed by current-controlled resistive switching induced by self-heating in polycrystalline $BaCo_{1-x}Ni_xS_{2-y}$ (nominal x=0.1 and y=0.2). These were due to the steep metal to insulator transition upon heating followed by the activated behavior of the resistivity above the transition. The major role of Joule heating in switching is supported by the absence of nonlinearity in the current as function of voltage, I(V), obtained in pulsed measurements, in the range of electric fields relevant to d.c. measurements. The voltage-controlled negative differential resistance around the threshold for switching was explained by a simple model of self-heating. The main difficulty in modeling I(V) from the samples resistance as function of temperature R(T) was the progressive increase of R(T), and to a lesser extend the decrease of the resistance jumps at the transitions, caused by the damage induced by cycling through the transitions by heating or self-heating. This was dealt with by following systematically R(T) over many cycles and by using the data of R(T) in the heating cycle closest to that of the self-heating one.

Most of the known first-order metal-to-insulator (M→I) transitions occur upon *cooling* from a high-T metal to a low-T insulator. M→I transitions in the opposite direction are unusual. A first-order transition from a low-T monoclinic paramagnetic metal to a high-T tetragonal antiferromagnetic insulator is exhibited by the layered compound $BaCo_{1-x}Ni_xS_{2-y}$ ($0.05<x<0.2$ and $0.05<y<0.2$)[1-3] and the related stoichiometric $BaCo_{0.9}Ni_{0.1}S_{2-y}Se_y$ series.[4] That the metallic phase, of lower symmetry structure,[2] has a lower entropy than the insulating, magnetically ordered phase seems counter-intuitive and puzzling. As shown below, this unusual M→I transition gives rise to electrical and thermal instability of an uncommon type.

Self-heating of a conductor under high enough d.c. voltages leads to I-V characteristics having regions of negative-differential-resistance (NDR). The trace of NDR is S-shaped (current controlled-CC) for dR/dT<0 and N-shaped (voltage controlled-VC) for dR/dT>0. In the case of CC-NDR, a maximal voltage is reached from which the sample switches toward a steady-state at lower voltage and higher current.[5] In the case of VC-NDR, a maximal current is reached from which the sample switches towards a steady state at higher voltage and lower current.[6] The slope of the I(V) line traced upon switching depends upon the circuit (the load line). In the case of a first-order insulator to metal (I→M) transition upon heating, dR/dT is very large and negative and CC-NDR is easily attained. In the case of a first-order metal-insulator (M→I) transition upon heating dR/dT is very large and positive and VC-NDR is easily attained. An even more interesting case is when dR/dT>>0 at the M→I transition is followed by a region of dR/dT<0 in the second (non-metallic) phase, that is, when R(T) above the transition is activated.

For the last fifty years the prototype system exhibiting a first-order transition from a low-T insulator to a high-T metal was VO$_2$ (T$_{I \to M}$ =340 K) and its popularity continues to grow due to its technological potential and interesting physics.[7,8] In VO$_2$ the d.c. I-V characteristics below T$_{I \to M}$ exhibits CC-NDR and associated instabilities driven by thermal and thermo-electric effects.[9,10] The role of Joule heating in the voltage-induced switching in VO$_2$ was debated for decades; it seems that now it is accepted as predominant. The advantages of this system are T$_{I \to M}$>RT (room temperature) and availability of high quality thin films, elastic beams and nano-beams that enable reproducibility of results over many cycles of cooling-heating through the transition, in spite of the large volume changes involved.

In our previous work on BaCo$_{1-x}$Ni$_x$S$_{2-y}$ (BaCoNiS)[11] we investigated samples with nominal x=0.1 and y=0.2. For these samples T$_{M \to I}$ is around 200K with a huge hysteresis of $\Delta$T~ 40 K; the resistance jump in fresh samples is 10<$\Delta$R/R$_m$<100 and the volume change is $\Delta V/V$~0.01 (comparable with that of VO$_2$ in the I→M transition). In polycrystalline samples this relatively large $\Delta V/V$ causes significant damage during temperature cycling. This is accompanied by major increase of the resistance of the samples upon each cycle and to a lesser extent a decrease in the height of the resistance jump. Nevertheless, the samples do not disintegrate even after many cooling-heating cycles. Also, the temperature width of the hysteresis loop and the local activation energy of the resistivity above the transition are preserved.

The steep dR/dT>0 at the transition with the potential for VC-NDR, followed by dR/dT<0 (activated transport) above the transition with potential for CC-NDR (both under self-heating) motivated us to investigate the nonlinear conductivity of polycrystalline BaCoNiS, in the absence of more robust samples.

The details of the sample preparation and additional properties can be found in our earlier publication.[11] The ohmic resistivity ($\rho$) was measured by the standard four-probe method in an automatic system. The pulsed and d.c. I-V measurements on samples with four-probes were carried out manually in a separate sample holder, as described in Ref. 12. In the following d$_{23}$, V$_{23}$, and R$_{23}$ denote the distance, voltage drop and resistance between the voltage probes, respectively.

A very thin BaCoNiS sample (labeled P) was prepared for the pulsed measurements which were limited to currents of 100 mA. Its dimensions were d$_{23}$=0.122 cm and cross-section area $A$=0.024×0.086 cm$^2$. Fig. 1(a) shows the pulsed I-V characteristics of sample P at various temperatures in the heating branch of the M→I transition and above it. The width of the pulses was 0.5 msec. The traces corresponding to the metallic state (T≤220 K) are linear up to current densities of at least 40 A/cm$^2$; above the transition the traces become nonlinear beyond ~20 A/cm$^2$, but the nonlinearity decreases with increasing T (decreasing $\rho$(T)). R(T) (and $\rho$(T)) shown in Fig.1(b) represent the inverse slopes of the lines in the linear regime of I-V. The weak nonlinearity obtained by using short pulses is probably of electronic origin (such as E-field enhanced hopping) and not due to self-heating.

$\rho(T)$ and d.c. I-V characteristics close to $T_{M \rightarrow I}$ were measured on several samples. Following the upward jump of $\rho(T)$ upon each heating (or self-heating)-cooling cycle over a number of cycles n, showed that $\rho(n)$ at fixed temperature was changing in a systematic manner. The switching loops of all samples were essentially similar, however the most detailed $\rho(n)$ results and simplest I-V loops were obtained from a sample labeled A. Its dimensions were $d_{23}$=0.30 cm and $A$=0.17× 0.42 cm$^2$.

Fig. 2(a) shows $\rho(T)$ loops for sample A during the cooling and heating cycles through the transitions in the automatic system, or measured manually during cooling from RT or heating and cooling prior and after the I-V measurements. For the sake of clarity we show only two traces of $\rho(T)$, out of the eight, carried out before and after I-V measurements. The resistance jump at set temperature $T_o$ in each I-V measurement appears in this figure as a vertical line that connects the ohmic resistivities before and after switching. This figure exhibits the difference between the phase transition due to external heating and that induced by self-heating. In the cooling branch, the resistance increases with decreasing temperature until $T_{I \rightarrow M}$ is reached and then drops abruptly. In the I-V measurements the sample's resistivity starts the cycle at $\rho(T_o)$ in the lower (metallic) branch and ends in the high-T (non-metallic) branch, but upon slow cooling it drops gradually. This indicates that the high-temperature state of the material after switching is different from the previous case – probably it is inhomogeneous, with various portions of the sample undergoing the I→M transition at different temperatures.

Fig. 2(b) displays the "penalty" of this sample for cycling through the structural phase transition. It shows $\rho$ at 110 K upon heating in the automatic system and upon cooling in the manual system from RT or upon heating in this system prior to the I-V measurements towards the M→I transition; it also shows $\rho_{max}$ - the value of $\rho$ before the I→M transition upon cooling in the automatic and in the manual systems. The values of n are integers for $\rho_{max}$ and half-integers for $\rho(110K)$. Over 14 cycles $\rho(110 K)$ increases by about an order of magnitude while $\rho_{max}$ increases by about a factor of 6. The height of the resistance jump decreases accordingly. A SEM micrograph of sample A at the end of the measurements (in the inset of 2(b)) shows deep gorges, probably created by the repeated cycling through the transitions.

Fig. 3(a) shows the d.c. I-V characteristics of sample A at three consecutively rising set temperatures $T_o$=195, 200 and 210 K within the thermal hysteresis range. The maximal currents and voltages in this Figure correspond to current densities and electric fields in their lowest ranges in Fig. 1(a). At low currents I(V) is linear up to a threshold current which decreases with rising $T_o$. The slopes of the lines also decrease with increasing $T_o$. Part of the corresponding increase in R is intrinsic and part is due to the damage added in the preceding cycle. Above the threshold current for onset of NDR, the sample switches tracing a straight line (the load line) towards a maximum voltage and then returns along the same line to an intermediate steady state. ***The back and forth switching are not isothermal***; a diode located close to the sample in the sample-holder showed cooling during the voltage increase and heating upon return, in both cases by about 20-30 K. The cooling of the surroundings was due to the heat absorbed by the sample during the M→I transition while heating was caused by the Joule heat emitted by the sample in the high resistance state. Fig. 3(b) shows the corresponding $R_{23}(V)=V_{23}/I$ obtained from Fig. 3(a) for $T_o=195$ K. The inset represents R(T) obtained from the closest $\rho(T)$ thermal loop (in Fig. 2 (a)) below the I-V measurements, fitted to R(195 K) prior to switching. Since this plot does not account for the degradation of the resistance jump due to the intermediate switching events, the peak resistance ($R_{max}$) may be overestimated by up to 20%. Fig. 3(b) shows that $R(V_{23})$ jumps from a constant value, to a value close to that of the peak in the inset and then jumps back to a minimum close to the R(RT). From this point on I(V) is stable and $R(V_{23})$ approaches steady state, increasing with decreasing $V_{23}$, eventually reaching a value close to the estimated R(195K) in the upper branch of R(T). We used the simplest model of self-heating by equating the power dissipated in the sample ($IV_{23}$) with that in Newton's law of cooling, $IV_{23}=\alpha(T-T_o)$, as in Ref. [13], assuming constant $T_o$. The dashed line in Fig. 3(a) is calculated for $T_o=195$ K and $\alpha=0.016$ W/K. Similar curves were obtained also for the higher $T_o$ showing that the threshold power for the onset of NDR decreases with increasing $T_o$. The calculated line shows also CC-NDR behavior beyond its minimum. The CC-NDR range is not applicable as is, since the process is ***not isothermal***; however the reduction of $T_o$ by 20-30 K (as observed in the experiment) shifts the line towards much larger values of $V_{23}$. This is illustrated by the additional line calculated for $T_o=170$ K and $\alpha=0.02$. Such a shifted I(V) line with CC-NDR causes the backward switching.

In conclusion, we have shown here unusual VC-NDR resistive switching followed by CC-NDR resistive switching induced by self-heating in $BaCo_{1-x}Ni_xS_{2-y}$ (nominal x=0.1 and y=0.2). These were due to the steep M→I transition upon heating followed by the activated behavior of the resistivity above the transition. The major role of Joule heating in switching is supported by the absence, or by the presence of only minor I(V) nonlinearity obtained in pulsed measurements. The back and forth I(V) switching is accompanied by a cooling-heating pulse that deserves further investigation.

Acknowledgements

The SEM work of Mrs. Ayelet Graff is gratefully acknowledged.

Figure 1. (a) Pulsed I-V measurements on BaCoNiS-P sample. Note that for T≤ 220 K ( the metallic state) the I(V) plots are linear up to current densities of at least 40 A/cm$^2$ (scale at right). Above the transition I(V) plots are nonlinear beyond ~20 A/cm$^2$, but the nonlinearity decreases with decreasing $\rho(T)$ (see the trace for T=255 K). (b) R(T) (scale at left) or $\rho(T)$ (scale at right) upon heating.

Figure 2. (a) $\rho(T)$ loops for BaCoNiS-A sample. Solid lines represent measurements in the automatic system (labeled "auto"), full circles represent data measured upon cooling from RT (labeled "man") and full diamonds – low current data before and after I-V measurements. For clarity only two out of eight traces of diamonds, the first and the last (I-V(1) and I-V(8)), are included here. (b) The effect of damage on the resistivity of sample A. The upper trace represents $\rho_{max}$ ($\rho$ at the I→M transition upon cooling) and the lower trace – $\rho$( 110K) upon heating in the automatic system, upon cooling in the manual

system from RT and upon heating prior to the I-V measurements (labeled I-V). Inset: SEM micrograph of sample A after the last $\rho(T)$ loop shown in (a).

Figure 3. (a) I-V characteristics of sample A at set temperatures $T_o$= 195 K, 200 K and 210 K. The calculated $I(V_{23})$ traces from the heating branch are shown only for the upper loop ($T_o$=195 K-dashed and $T_o$=170 K-dot-dashed). (b) $R_{23}(T)$ obtained from the experimental data in (a). Inset: $R_{23}(T)$ from Fig 2(a) fitted to $R_{23}(195\ K)$, used for the calculated curves. $R_{23}$ and LL denote the resistance between voltage probes and the load line, respectively.

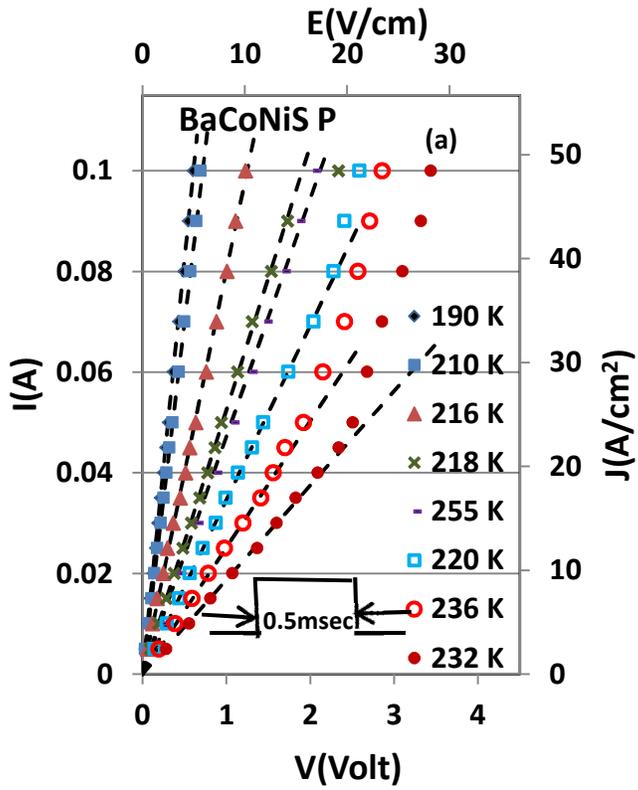
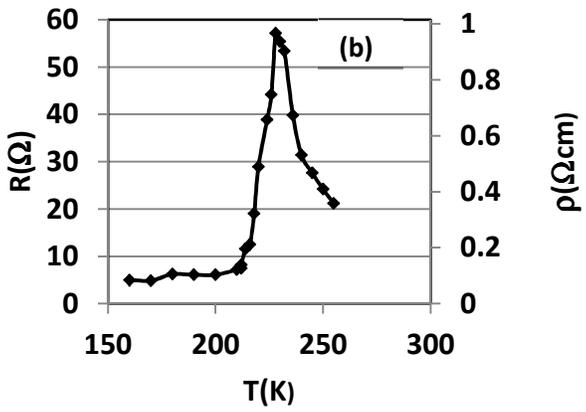

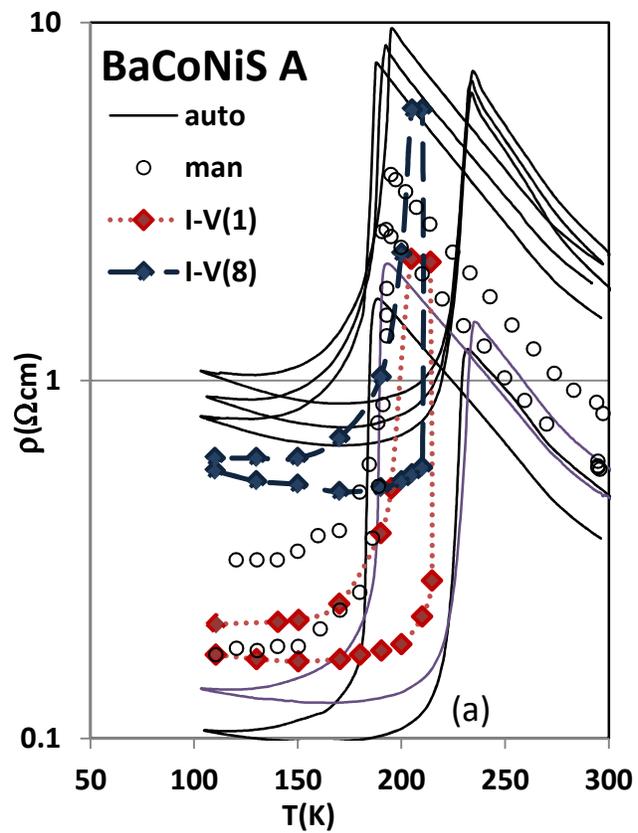

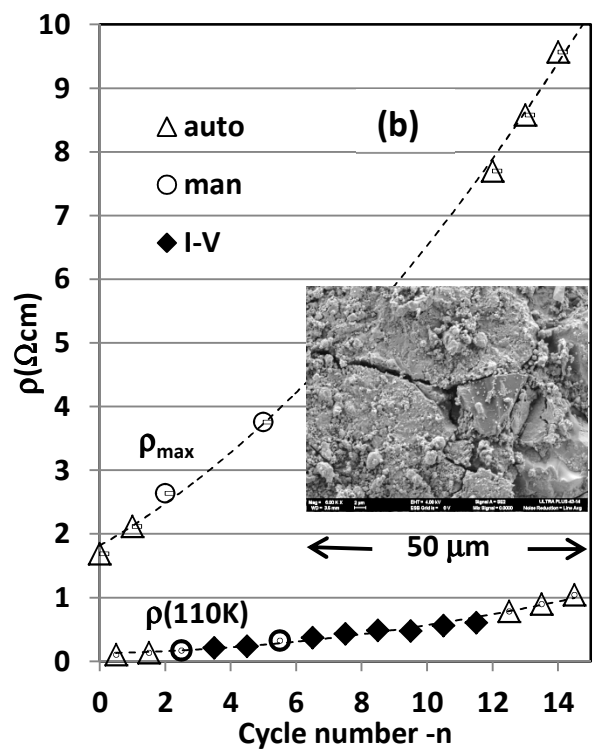

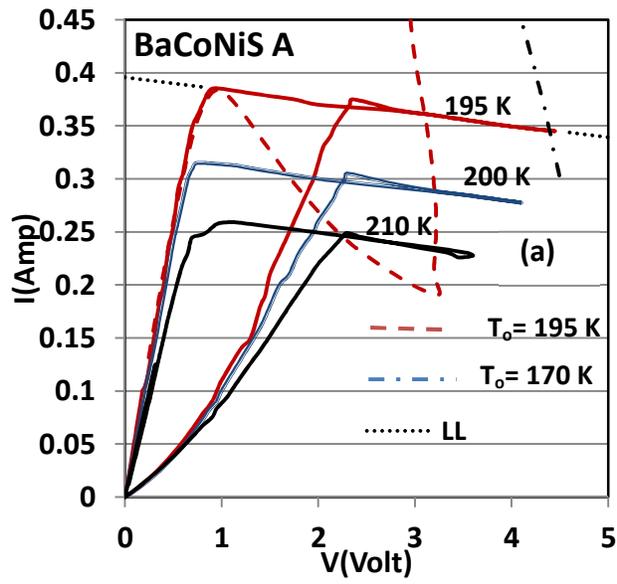

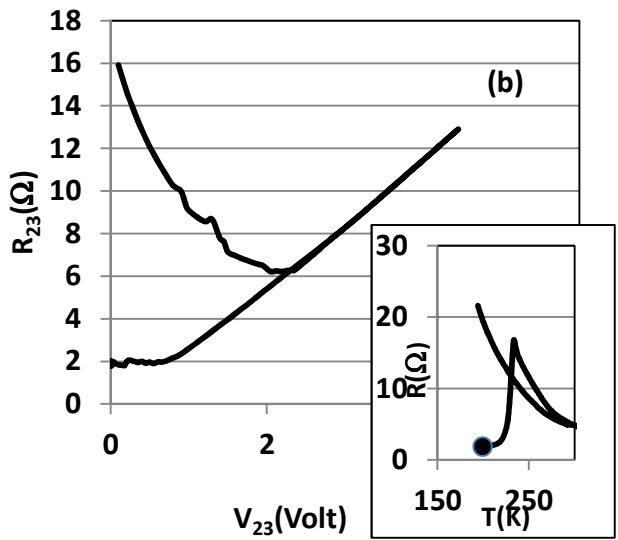